\documentclass[%
 aip,
 amsmath,amssymb,
 reprint,%
]{revtex4-1}

\usepackage{graphicx,subfigure}
\usepackage{dcolumn}
\usepackage{bm}
\usepackage{float}
\usepackage{xcolor}
\usepackage{placeins}
\usepackage{listings}
\usepackage{siunitx} 
\usepackage[version=4]{mhchem} 
\usepackage{xpatch}

\makeatletter
\def\@email#1#2{%
 \endgroup
 \patchcmd{\titleblock@produce}
  {\frontmatter@RRAPformat}
  {\frontmatter@RRAPformat{\produce@RRAP{*#1\href{mailto:#2}{#2}}}\frontmatter@RRAPformat}
  {}{}
}%
\makeatother

\newcommand{\matr}[1]{\bm{#1}} 
\newcommand{\vect}[1]{\bm{#1}}

\definecolor{mygreen}{rgb}{0,0.6,0}
\definecolor{mygray}{rgb}{0.5,0.5,0.5}
\definecolor{mymauve}{rgb}{0.58,0,0.82}
\PassOptionsToPackage{hyphens}{url}\usepackage{hyperref}

\newlength{\figurewidth}
\begin{document}

\title{Updates to the DScribe Library: New Descriptors and Derivatives}


\author{Jarno Laakso}
\affiliation{%
 Department of Applied Physics, Aalto University, P.O. Box 11100, 00076 Aalto, Finland
}%

\author{Lauri Himanen}
\affiliation{%
 Department of Applied Physics, Aalto University, P.O. Box 11100, 00076 Aalto, Finland
}%

\author{Henrietta Homm}
\affiliation{%
 Department of Applied Physics, Aalto University, P.O. Box 11100, 00076 Aalto, Finland
}%

\author{Eiaki V. Morooka}
\affiliation{%
 Department of Applied Physics, Aalto University, P.O. Box 11100, 00076 Aalto, Finland
}%

\author{Marc O. J. J\"ager}
\affiliation{%
 Department of Applied Physics, Aalto University, P.O. Box 11100, 00076 Aalto, Finland
}%

\author{Milica Todorović}
\affiliation{Department of Mechanical and Materials Engineering, University of Turku, FI-20014 Turku, Finland}

\author{Patrick Rinke}
\affiliation{%
 Department of Applied Physics, Aalto University, P.O. Box 11100, 00076 Aalto, Finland
}%

\date{\today}

\begin{abstract}
We present an update of the DScribe package, a Python library for atomistic descriptors. The update extends DScribe's descriptor selection with the Valle-Oganov materials fingerprint and provides descriptor derivatives to enable more advanced machine learning tasks, such as force prediction and structure optimization. For all descriptors, numeric derivatives are now available in DSribe. For the many-body tensor representation (MBTR) and the Smooth Overlap of Atomic Positions (SOAP), we have also implemented analytic derivatives. We demonstrate the effectiveness of the descriptor derivatives for machine learning models of Cu clusters and perovskite alloys. 
\end{abstract}

\maketitle

\section{\label{sec:introduction}Introduction}
DScribe is a software library that provides atomistic descriptors to researchers in the natural sciences and engineering.\cite{himanen2020dscribe} Descriptors represent the atomic structure of molecules, nanostructures and materials in machine-readable format. To facilitate machine learning (ML), descriptors should be invariant under transformations that conserve physical quantities, such as translations, mirroring, rotations and atomic permutations.\cite{Langer2022} Descriptors are also a powerful tool for defining distance metrics between atomic structures, which is helpful in many ML tasks, such as clustering or kernel-based regression. While ML is one of the main applications of descriptors, their usefulness is not limited to that. They can also be utilized, for example, in similarity analysis of atomic structures or visualization of atomistic data. In this article we present new features that we have added to DScribe, including a new descriptor and the capability to calculate descriptor derivatives with respect to atomic positions.

At the time of its publication in 2019, DScribe included six descriptors: the Coulomb matrix,\cite{rupp2012fast} the sine matrix,\cite{faber2015crystal} the Ewald sum matrix,\cite{faber2015crystal} the Many-Body Tensor Representation (MBTR),\cite{huo2017unified} the Atom-centered Symmetry Functions (ACSF) \cite{behler2011atom-centered} and the Smooth Overlap of Atomic Positions (SOAP).\cite{bartok2013on} The first four descriptors in this list are global descriptors and the remaining two local. DScribe made these descriptors available to a wide community and facilitated a variety of machine learning applications including property prediction,\cite{fung2021benchmarking, zhou2022machine, pihlajamaki2020monte, rahaman2020deep, sun2022machine, hirai2022machine-learning} global structure search,\cite{lourenco2021taking} and data analysis and visualization.\cite{sun2021ab, cheng2020mapping, monserrat2020liquid} All DScribe descriptors can output the representations in vector form, which makes them compatible with a multitude of existing ML methods and algorithms. Thus far, they have been employed, for example, with linear regression models,\cite{sun2022machine, hirai2022machine-learning, zhou2022machine} neural networks,\cite{fung2021benchmarking, zhou2022machine, rahaman2020deep, sun2022machine} random forests,\cite{zhou2022machine, sun2022machine} and Gaussian processes.\cite{lourenco2021taking}

In this work, we present a new descriptor we recently added to DScribe. It is based on a structural fingerprint proposed by Valle and Oganov for similarity analysis of crystal structures.\cite{valle2010crystal} Since its formulation, it has been adopted for other applications. Bisbo \textit{et al}. used the  Valle-Oganov fingerprint to represent atomic structures in their global structure optimization algorithm.  \cite{bisbo2020efficient} Arrigoni \textit{et al}. combined it with principal component analysis (PCA) for dimensionality reduction to facilitate data analysis.\cite{arrigoni2021evolutionary} According to its original definition, the Valle-Oganov fingerprint is constructed from the interatomic distances in an atomic structure. It greatly resembles the $k$=2-term of the MBTR descriptor, but unlike MBTR, it was specifically tailored for periodic systems. The more specific use-case reduces the number of user-defined parameters, which makes the Valle-Oganov descriptor easier to use than MBTR. Other studies have extended on the Valle-Oganov fingerprint by adding an angular term.\cite{bisbo2022global} We also included such a higher-order term in our implementation of the Valle-Oganov descriptor, but allow the user to decide whether to use it or not.

Structural descriptors have facilitated quick and accurate property prediction of molecules and materials using ML.\cite{faber2015crystal, stuke2019chemical, jiang2021topological} Many useful properties are derivatives of other quantities, which means that the same ML models can be used to predict multiple quantities. For atomic structures, gradients of the energy give access to atomic forces. Differentiating an ML model for the energy with respect to atomic positions would therefore also provide force predictions. This is the working principle of ML potentials that are increasingly employed in simulating the dynamics of atomic systems.\cite{behler2007generalized, bartok2010gaussian, schutt2018schnet} Having access to the derivatives of an ML model also helps with optimization of the predicted property. Energy minimization, for example, is one of the most common tasks in computational chemistry and physics. Using an ML model to relax atomic positions by optimizing the surrogate potential energy surface instead of using a more expensive method such as density functional theory (DFT) saves computational resources. Descriptor derivatives enable these ML tasks but they have not been generally available to date. In this article, we present our work on implementing descriptor derivatives in DScribe, with the goal to make them more accessible to the research community.

Descriptor derivatives can be calculated either analytically or numerically. Numerical derivatives are easy to implement and can be transferred to all descriptors in the library. Their disadvantage is the increased numerical cost and potential discretization errors. The number of descriptor evaluations required to calculate the numerical derivatives with respect to all atomic coordinates, for example, scales linearly with the system size. Analytical derivatives do not suffer from limited accuracy, and the computational time for calculating all the derivatives of a descriptor analytically is usually comparable to a single descriptor calculation. The downside of analytical derivatives is that they need to be implemented separately for every descriptor, which can be very tedious. Here, we have implemented analytical derivatives for SOAP and MBTR and numerical derivatives for all descriptors. We demonstrate the effectiveness of the SOAP derivatives by employing them  in a neural network model to fit a simple ML potential. We also showcase the application of MBTR derivatives in a structure optimization task.

This article is organized as follows: Section \ref{sec:methods} presents a description of the numerical and analytical descriptor derivatives, as well as the Valle-Oganov descriptor. In Section \ref{sec:software-structure}, we elaborate on the practical implementation of the new features in DScribe, and provide a guide on their usage. Section \ref{sec:results} presents the results for tests that we conducted to make sure that the descriptor derivatives function properly and showcase their usefulness with two demonstrations. Finally, in Section \ref{sec:conclusions}, we conclude our work.

\section{\label{sec:methods}Methods}
In this section we detail the computational methods behind the features that we added to DScribe. We describe our approach for computing numerical descriptor derivatives with the central difference method and detail our implementations of the analytic SOAP and MBTR derivatives. We then briefly present two ML tasks that demonstrate the effectiveness of our implementation. We close the section by introducing the additional descriptor that is based on the structural fingerprint by Valle and Oganov, and showing how we implemented it with small changes to the existing MBTR implementation.

\subsection{\label{sec:derivatives}Descriptor derivatives}

\subsubsection{Numerical derivatives}
The derivatives of any atomic representation $F(\matr{R})$ with respect to the atomic positions can be approximated using the central finite difference method.
\begin{align}
    \frac{\partial F(\matr{R})}{\partial \matr{R}_{ij}} \approx \frac{F(\matr{R}+\frac{h}{2}\matr{E}^{ij})-F(\matr{R}-\frac{h}{2}\matr{E}^{ij})}{h},
\end{align}
where $\matr{E}^{ij}$ is a single-entry matrix:
\begin{align}
    (\matr{E}^{ij})_{ab} = \begin{cases}
    1, &\text{when } a=i \text{ and } b=j\\ 
    0, &\text{otherwise}
    \end{cases}
\end{align}
and $h$ quantifies the magnitude of the atomic displacement. The error made by the central difference approximation is proportional to $h^2,$ which means that the accuracy of the numerical derivatives improves quickly with smaller values of $h.$ In practice, however, the limited floating-point accuracy makes the derivatives unstable when $h$ is too small. In order to determine the optimal value of $h$ for our implementation, we conducted a test comparing the numerical derivatives computed with different $h$ values to the analytical derivatives that we derived for MBTR and SOAP descriptors. We calculated the relative error between the numerical and analytical derivatives using the mean relative percentage difference
\begin{align}
    \text{MRPD} = \frac{2}{3NM}\sum_{i,j,k}\left|\frac{\partial_{ij}^{\text{a}}F_k-\partial_{ij}^{\text{n}}F_k}{|\partial_{ij}^{\text{a}}F_k|+|\partial_{ij}^{\text{n}}F_k|}\right|,
\end{align}
where $i$ iterates over the atoms in an atomic structure, $j$ runs over the three Cartesian coordinates, and $k$ runs over the components of the representation vector. $N$ is the number of atoms in the structure and $M$ the number of components in the representation vector. $\partial_{ij}^{\text{a}}$ and $\partial_{ij}^{\text{n}}$ are the analytical and numerical derivatives with respect to coordinate $\matr{R}_{ij}$, respectively.

\subsubsection{MBTR derivatives}
In DScribe three MBTR terms are implemented, each corresponding to atomic motifs of a different size. The $k=1$ term encodes single atoms, the $k=2$ term atomic pairs, and the $k=3$ term atom triples. Each term is defined as
\begin{align}
    F(x) = N\sum_i w^i d^i(x),\label{eq:mbtr}
\end{align}
where the sum runs over the motifs in an atomic structure. $N$ is a normalization term, $w^i$ is the weight related to motif $i$ and $d^i(x)$ is a distribution function. DScribe uses the Gaussian distribution:
\begin{align}
    d^i(x) = \frac{1}{\sigma \sqrt{2\pi}}\exp\left(-\frac{(x-g^i)^2}{2\sigma^2}\right),\label{eq:mbtr_gaussian}
\end{align}
where $\sigma$ is the standard deviation of the distribution and $g^i$ is a function that maps atomic motif $i$ to a scalar value. Assuming that $N$ is independent of the atomic positions, the gradient of the representation function with respect to the coordinates of an atom is
\begin{align}
\nabla F(x) = N\sum_i d^i(x)\left[\nabla w^i + w^i \frac{1}{\sigma^2} (x-g^i) \nabla g^i\right].
\end{align}
$\nabla w^i$ and $\nabla g^i$ depend on the choice of the weighting and geometry functions. The assumption of $N$ being independent of atomic positions is not true for the normalization option \texttt{l2\_each} that normalizes the L2-norm of the representation vector to one, and analytical derivatives are currently not available for it. For a detailed derivation of the MBTR gradients with the different options for $\nabla w^i$ and $\nabla g^i$ we refer to Section S2.B of the Supplementary Material (SM).

To demonstrate the analytical MBTR derivatives, we used them in geometry optimization of perovskite materials. We fitted an energy ML model that combines the MBTR with kernel ridge regression (KRR) for a \ce{CsPb(Cl/Br)3} data set. We generated the data set and used the same ML model for an earlier study that contains detailed information on model fitting and structure optimization.\cite{laakso2022compositional} For this demonstration, we optimized the atomic positions of 25 perovskite structures using the ML model and compared the obtained energies and geometries to DFT relaxation results.

\subsubsection{SOAP derivatives}
\begin{figure}
    \includegraphics[width=\figurewidth]{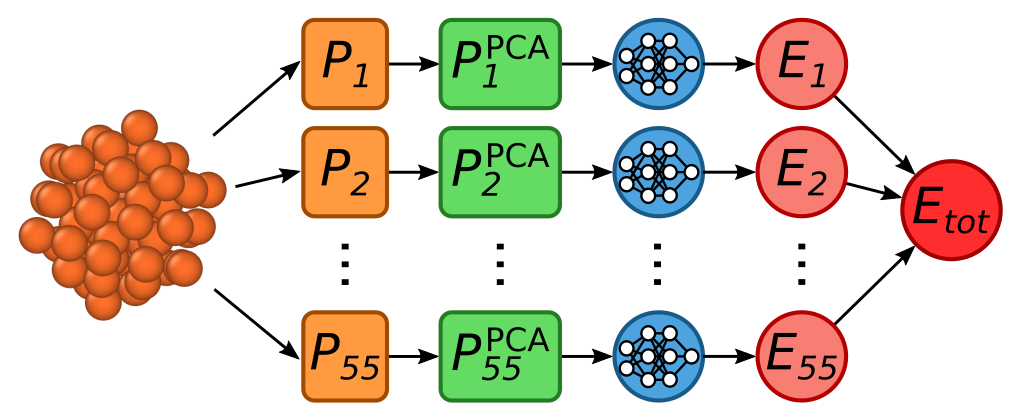}
    \caption{The ML model for Cu cluster energy prediction. All 55 atomic environments of a Cu cluster are first represented in vector form using SOAP ($P_i$). Then the dimensionality of these vectors is reduced with PCA. The PCA-reduced SOAP vectors $P_i^{\text{PCA}}$ are mapped to atomic energies $E_i$ with a feedforward neural network with two hidden layers. Finally, $E_i$ are summed together to obtain the total potential energy $E_{\text{tot}}$ of the Cu cluster.
    }
    \label{fig:fig-1}
\end{figure}
The SOAP descriptor represents local atomic environments in a rotationally invariant way by expanding Gaussian smeared atomic densities in a basis of spherical harmonics and radial basis functions. In DScribe, SOAP outputs a vector of partial power spectra \cite{de2016comparing} $\bm{p}$, where the individual components are defined as
\begin{align}
p^{Z_1,Z_2}_{nn'l}
&= \pi \sqrt{\frac{8}{2l+1}} \sum_m \left(c_{nlm}^{Z_1}\right)^*c^{Z_2}_{n'lm}\label{eq:power_spectrum_complex_1}\\
&= \pi \sqrt{\frac{8}{2l+1}} \sum_m \left(\sum_j^{\lvert Z_1 \rvert} c^j_{nlm}\right)\left(\sum_k^{\lvert Z_2 \rvert} c^k_{n'lm}\right).  \label{eq:power_spectrum_complex_2}
\end{align}
The complex conjugation in \ref{eq:power_spectrum_complex_1} can be omitted, because DScribe uses real spherical harmonics. The summations for
$j$ and $k$ run over atoms in the environment with the atomic numbers $Z_1$ and $Z_2,$ respectively. Coefficients $c^i_{nlm}$ are defined as
\begin{align}
c^i_{nlm}
&= \iiint_{\mathcal{R}^3}\mathrm{d}V g_{nl}(r)\rho(\bm{r}, \bm{R}_i)Y_{lm}(\theta, \phi), \label{eq:coefficient_atom}
\end{align}
where $\rho(\bm{r}, \bm{R}_i)$ is the Gaussian smeared atomic density, $g_{nl}(r)$ a radial basis function and $Y_{lm}(\theta, \phi)$ a spherical harmonic. The vector $\bm{p}$ consists of elements $p^{Z_1,Z_2}_{nn'l}$ for all unique atomic number pairs $(Z_1,Z_2),$ and unique combinations of radial and spherical basis functions $(n,n',l).$

The gradient of $p^{Z_1,Z_2}_{nn'l}$ with respect to the coordinates of an atom is
\begin{align}
\nabla p^{Z_1,Z_2}_{nn'l}
= \pi \sqrt{\frac{8}{2l+1}} &\sum_m \left[ \left( \sum_j^{\lvert Z_1 \rvert} \nabla c^j_{nlm} \right) \sum_k^{\lvert Z_2 \rvert} c^k_{n'lm}\right.\nonumber\\
&+ \left.\sum_j^{\lvert Z_1 \rvert} c^j_{nlm} \left( \sum_k^{\lvert Z_2 \rvert}\nabla c^k_{n'lm} \right) \right]. \label{eq:pnnl_derivative}
\end{align}
The final derivative equation depends on the choice of radial basis function. For now, we have implemented the analytical derivatives for spherical primitive Gaussian type orbitals in the non-periodic case. For polynomial radial basis functions and periodic systems, numerical differentiation is used instead. See section S3.B of SM for the full derivation of the SOAP gradients.

We tested the SOAP derivatives in an ML potential model that we trained for Cu clusters. We generated data for training and testing the ML potential by running a classical molecular dynamics simulation of a 55-atom Cu cluster at \SI{500}{K} for \SI{5.0}{ns} and uniformly picking 10 000 snapshots from the simulation. The simulation used a \SI{5.0}{fs} timestep, embedded atom method \cite{foiles1986embedded-atom-method} (EAM) for the Cu interactions, and Nose-Hoover thermostat \cite{hoover1985canonical} for the temperature control. It was performed using the LAMMPS simulation tool.\cite{thompson2022lammps}

We built an ML model that first represents all 55 atomic environments in vector form using SOAP, and then uses a feedforward neural network to map the atomic environments to atomic energies, which are summed to the total energy. To make model training easier, we decreased the dimensionality of the SOAP vectors with principal component analysis (PCA) following an earlier study by Zhou \textit{et al}.\cite{zhou2022machine} The full ML model architecture is depicted in Fig. \ref{fig:fig-1}.

To generate the SOAP vectors we used a radial cutoff of \SI{6.0}{\text{\AA}} and a basis of 8 radial basis functions and 6 spherical harmonics. These settings result in 252 dimensional SOAP vectors. We reduced the dimensionality of these vectors to 50 with PCA. The neural network that we used to map these reduced SOAP vectors to atomic energies had two hidden layers. Both hidden layers had 50 nodes that used the sigmoid activation function. The output layer used linear activation. We implemented the neural network using Keras \cite{chollet2015keras} and Tensorflow \cite{abadi2015tensorflow} Python packages. The weights of the network were optimized using the Adam algorithm.\cite{kingma2014adam}

We trained the ML potential on total energies of 8 000 Cu clusters. Then we assessed the quality of the fit by predicting the total energies of the remaining 2 000 clusters and compared the results to the EAM energies. Atomic forces are derivatives of the total energy with respect to the atomic positions. By combining our implementation of the SOAP derivatives with the gradients of the neural network, we were able to use the ML model to predict atomic forces. We used the ML model to predict the forces of all 2 000 test set clusters and compared the results to the EAM forces.

\subsection{\label{sec:valle-oganov}Valle-Oganov descriptor}
In their article Valle and Oganov defined an atomic structure representation for periodic systems.\cite{valle2010crystal} The full representation is a matrix in pairs of elements $(A,B):$  
\begin{align}
\label{eq:VO}
    F_{AB}(x)=\sum_{A_i B_j}\frac{d^{i,j}(x)}{4\pi r_{ij}^2 (N_A N_B/V)} - 1.
\end{align}
The index $i$ runs over all atoms of type $A$ and $j$ over all atoms of type $B.$ $r_{ij}$ is the distance between atoms $i$ and $j, $ $V$ the volume of the cell and  $N_A$ and $N_B$ the number of atoms of each type. $d^{i,j}(x)$ is the Gaussian distribution of equation \ref{eq:mbtr_gaussian} with the geometry function set to be the distance between the atoms $g^{i,j}=r_{ij}.$

Closer inspection of equation~\ref{eq:VO} reveals close similarity to the MBTR, which is already implemented in DScribe. The only fundamental difference between the two representation is the constant term $-1$ in the Valle-Oganov representation. The constant term, however, is not significant in most use-cases, as it does not affect the distances between the vectors. Furthermore, if the constant term is needed for some application, it can be added to the representation vectors afterwards. We therefore decided to omit the constant term $-1$ from our implementation. Now $F_{AB}(x)$ can be cast into the general MBTR formalism defined in equation \ref{eq:mbtr} by setting the weighting function to
\begin{align}
    w^{i,j} = \frac{1}{r_{ij}^2}
\end{align}
and the normalization term to
\begin{align}
    N_{AB} = \frac{V}{4\pi N_A N_B}.
\end{align}
The Valle-Oganov descriptor implementation in DScribe therefore only requires the addition of new weighting and normalization options to the already existing MBTR implementation.

\begin{figure}
    \includegraphics[width=\figurewidth]{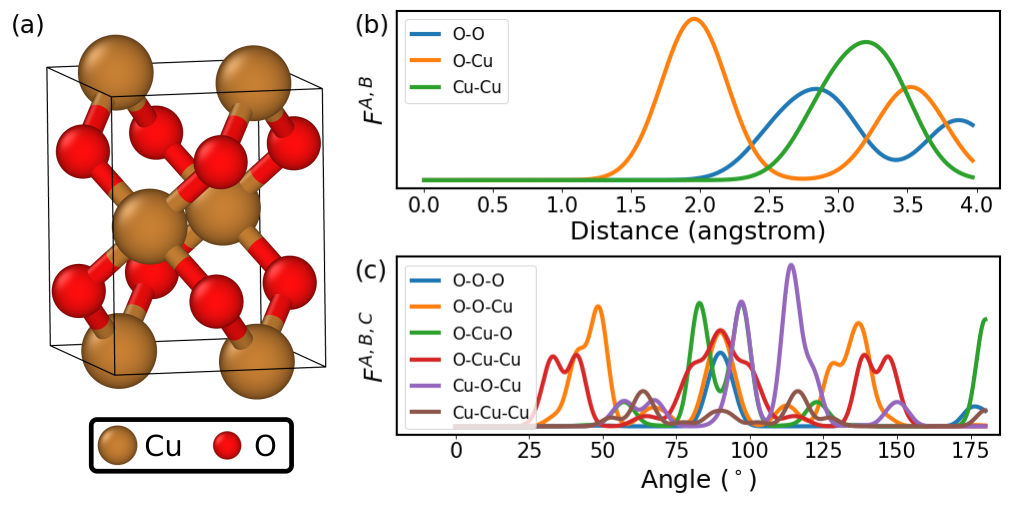}
    \caption{Valle-Oganov descriptor construction. (a) \ce{CuO} atomic structure. (b) Second-order and (c) third-order Valle-Oganov fingerprints of \ce{CuO}. The full Valle-Oganov representation vector is obtained by concatenating the elemental contributions.
    }
    \label{fig:fig-2}
\end{figure}
The original article by Valle and Oganov only considers interatomic distances, but the representation has been extended with an angular term by Bisbo and Hammer.\cite{bisbo2022global} We define a similar third-order term $F_{ABC}(x),$ again utilizing the MBTR formalism. Now, the sum in equation \ref{eq:mbtr} runs over atom triplets $(i,j,k)$ and the geometry function in equation \ref{eq:mbtr_gaussian} is the angle between the atoms $g^{i,j,k}=\theta_{ijk}.$ The normalization term is
\begin{align}
    N_{ABC} = \frac{V}{4\pi N_A N_B N_C}
\end{align}
and the weight function is
\begin{align}
    w^{i,j,k} = f_c(r_{ij}) f_c(r_{ik}),
\end{align}
where
\begin{equation}
    f_c(r)=1+\gamma\left(\frac{r}{r_{\text{cut}}}\right)^{\gamma+1}-(\gamma+1)\left(\frac{r}{r_{\text{cut}}}\right)^\gamma.
\end{equation}
The weight function and its derivative approach 0 at the cutoff distance $r_{\text{cut}}$ beyond which the atom triplets do not contribute to the representation. $\gamma$ controls the sharpness of the cutoff and in our implementation it has a default value of 2. 

Fig. \ref{fig:fig-2} visualizes the Valle-Oganov fingerprint of \ce{CuO}. The full representation vector is obtained by concatenating the different elemental contributions of the second and third-order terms. Although we use the MBTR framework for the Valle-Oganov descriptor, analytical derivatives have not yet been fully implemented for its normalization and weighting options, and numerical differentiation is used instead.

\section{\label{sec:software-structure}Software structure}
Python has solidified its position as a go-to language for several domains, including data science. In order to seamlessly integrate with these Python-based data science workflows, the main entry point for our software is a Python API through which the descriptors can be configured and used. Figure \ref{fig:interface} shows an example of the new DScribe interface for calculating derivatives.

\begin{figure}[h]
    \centering
\begin{lstlisting}[language=Python]
from ase.build import molecule
from dscribe.descriptors import SOAP

# Define the atomic systems
systems = [molecule("H2"), molecule("O2")]

# Setup the descriptor
soap = SOAP(
    species=["H", "O"],
    rbf="gto",
    rcut=3,
    nmax=10,
    lmax=8,
    sparse=True
)

# Calculate derivatives and descriptor features
derivatives, features = soap.derivatives(
    systems,
    positions=None,
    include=[0],
    exclude=None,
    method="auto",
    return_descriptor=True,
    n_jobs=2
)
\end{lstlisting}
    \caption{Example of creating derivatives with DScribe. All descriptors have the \texttt{derivatives}-function that can be used to retrieve both derivatives and descriptor features simultaneously. Only the first argument that specifies the used systems is required and the additional arguments can be used to further control the methodology, parallelization, and with respect to which atoms the positional derivatives are calculated for.}
    \label{fig:interface}
\end{figure}

DScribe works with atomic structures defined using the \texttt{ase.Atoms}-object from the \texttt{ase} package.\cite{ase-paper} These objects are easy to create from existing structure files or build with the utilities provided by \texttt{ase}. In addition to any descriptor-specific arguments, all descriptors accept the \texttt{sparse}-parameter that controls whether the created output is a dense or a sparse matrix. Especially in large systems where the interactions between atoms and centers of interest are very localized, sparsity provides memory and storage efficiency. Some ML algorithms can use sparse matrix formats directly, but it is also easy to restore the dense format for other algorithms. 

All descriptors implement the new \texttt{derivatives} method. The first argument accepts one or multiple atomic structures. The argument \texttt{positions} can be used to define the positions of interest for local descriptors. It defaults to using all individual atoms as centers and cannot be specified for global descriptors. The arguments \texttt{include} and \texttt{exclude} are used to control which atoms are considered in the derivative calculations. By using \texttt{method}, the user can explicitly change between analytical and numerical differentiation. The default value \texttt{auto} will automatically choose the analytical implementation if it is available, and the numerical one otherwise. Descriptors and their derivatives can be created simultaneously by setting \texttt{return\_descriptor=True} as this is often computationally favorable. Finally, descriptor calculations can be parallelized over several CPU cores using the \texttt{n\_jobs} parameter. This parallelization is based on evenly distributing the given atomic structures to different cores for data parallelism.

We have decided to retain as much structure in the derivative output as possible. This approach allows the user to better understand and access the different components, while it is still relatively easy to re-arrange the output into a lower-dimensional shape if needed. Generally, the output is a multidimensional array with shape \texttt{[n\_systems, n\_centers, n\_atoms, 3, n\_features]}. Here the dimension with \texttt{n\_systems} runs over the different atomic structures, \texttt{n\_centers} loops through the different centers of interest, \texttt{n\_atoms} loops through the atoms for which the derivatives were calculated, the second-to-last dimension with three components loops through the x, y, and z components, and the last dimension with \texttt{n\_features} loops through the different descriptor features. Global descriptors effectively have only one region of interest that covers the entire structure, meaning that \texttt{n\_centers = 1} and the corresponding dimension is not present. Similarly, when creating the derivatives only for one system, \texttt{n\_systems = 1}, and that dimension is left out.

As many of the descriptor calculations require significant CPU resources, many of the heavier calculations are internally handled by an underlying C++ implementation that is accessed through the Python interface. This hybrid approach is similar to many other common numerical Python packages such as \texttt{numpy} \cite{numpy} and \texttt{scipy}.\cite{scipy} The communication between Python and C++ is implemented using the \texttt{pybind11} \cite{pybind11} library.

The source code is structured using an object-oriented programming approach. Each descriptor is represented by its own class which inherits from a generic base class. When adhering to the base class interface, the subclasses can automatically take advantage of the functionality already defined in the base class - such as the numerical derivatives - in addition to ensuring that the user can expect each descriptor to have similar functionality. Each descriptor class is associated with a code test suite that is used to ensure the validity of the implementation. This test suite forms part of a continuous integration pipeline that is performed every time the source code is modified.

DScribe is distributed as a Python package and can be installed from the Python package index (PyPI) \cite{pypi} under the package name \texttt{dscribe}. Alternatively, the package can be installed using the \texttt{conda-forge} \cite{conda_forge} ecosystem where it is distributed with the same name. Access to the full source code is also provided through GitHub at \url{https://github.com/SINGROUP/dscribe}. The full documentation and several tutorials are available on the DScribe homepage  \url{https://singroup.github.io/dscribe/}.

\section{\label{sec:results}Results and Discussion}
This section showcases the outcomes of the tests conducted on the descriptor derivatives. First, we assessed the accuracy of our numerical derivative implementation by comparing it to analytical derivative values. We used the results of the test to determine the optimal $h$-value for DScribe's numerical derivative implementation. We then present MBTR derivatives for the perovskite structures and the SOAP derivative test for the Cu cluster.

\subsection{Numerical descriptor derivatives}
\begin{figure}
    \includegraphics[width=\figurewidth]{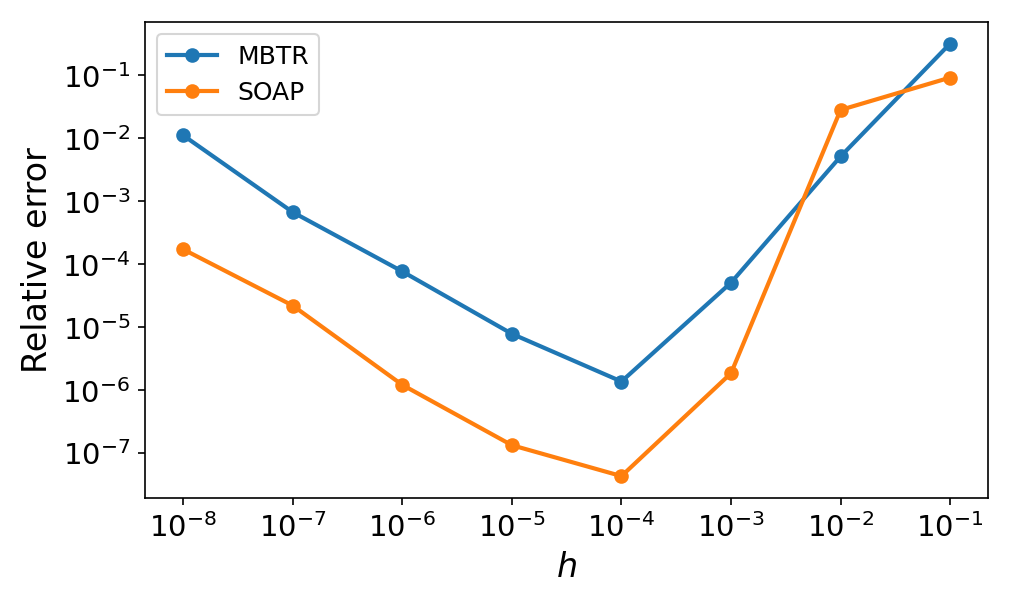}
    \caption{Relative error of numerical derivatives compared to analytical MBTR and SOAP derivatives with different central difference step sizes $h$.
    }
    \label{fig:fig-4}
\end{figure}
We tested the accuracy of our numerical descriptor derivative implementation by comparing it to analytical MBTR and SOAP derivatives. Fig. \ref{fig:fig-4} shows the relative error between the numerical and analytical derivatives for a water molecule. The errors are at their highest at the higher end of the tested $h-$range. With decreasing $h$ the relative error reduces due to the $h^2$ scaling of the central difference error. The MBTR and SOAP errors both reach their minima at $h=\SI{1e-4}{\text{\AA}},$ after which they start to increase again due to the limited floating-point accuracy. For both descriptors the relative error reaches $10^{-6}$ at its lowest. In practical applications, this is likely to be insignificant compared to errors related to model fitting or the underlying data. Based on the results of the test, we fixed $h=\SI{1e-4}{\text{\AA}}$ for the numerical derivatives of all descriptors in DScribe. The good agreement between the numerical and analytical derivatives shows that the numerical derivative implementation is highly accurate and that our analytical MBTR and SOAP derivatives are error-free.

\subsection{Perovskite structure optimization with MBTR}
\begin{figure}
    \includegraphics[width=\figurewidth]{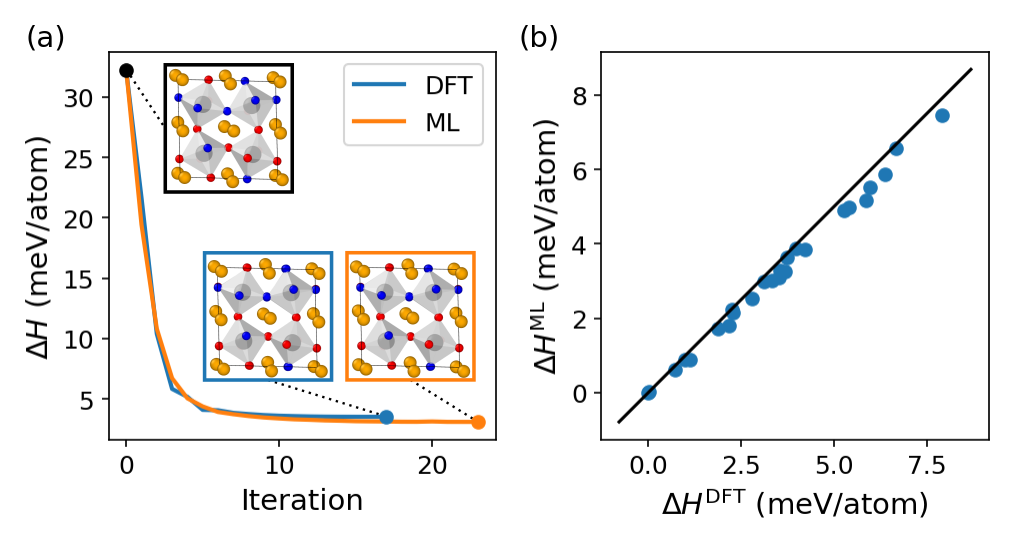}
    \caption{Results from perovskite structure optimization tests utilizing MBTR gradients. a) One perovskite structure optimized with both DFT and the ML model. b) Comparison of ML and DFT optimized energies of 25 perovskite structures.
    }
    \label{fig:fig-5}
\end{figure}
To showcase the analytical MBTR derivatives, we fitted an ML model that combines MBTR and KRR to predict energies of \ce{CsPb(Cl/Br)3} perovskite structures. To asses the accuracy of the fit, we used the model to predict the energies and forces of structure snapshots from DFT relaxation of 25 perovskite test structures. The mean absolute error (MAE) of energy predictions was only \SI{0.14}{meV/atom} while the force prediction MAE was \SI{16.7}{meV/\text{\AA}}.

We then used the derivatives of the trained ML model to relax the atomic positions of the same 25 \ce{CsPb(Cl/Br)3} test geometries. Fig. \ref{fig:fig-5}a shows an example of how the ML predicted energy of a perovskite structure decreases during ML relaxation, and how that compares to the DFT relaxation of the same structure. The DFT relaxation reaches the minimum structure in fewer iterations than the ML model, but the final energies are very close, deviating only by \SI{0.42}{meV/atom}. The final structures from the two relaxation methods are almost identical with the root-mean-square deviation (RMSD) of atomic positions between them being only \SI{0.016}{\text{\AA}}. Furthermore, relaxing the structure with the ML model is over four orders of magnitude faster than with DFT.

In Fig. \ref{fig:fig-5}b, we plotted the relaxed energies of all 25 perovskite test structures. For most structures, the DFT and ML optimized energies are nearly identical, although in some cases the energies differ by up to \SI{0.70}{meV/atom}. The mean absolute error between the two energies is \SI{0.26}{meV/atom} and the average RMSD of atomic positions is \SI{0.031}{\text{\AA}}. The good agreement in terms of both energy and structure demonstrates that an ML approach utilizing descriptor derivatives can effectively accelerate structure optimization computations.

\subsection{Cu cluster ML potential using SOAP}
\begin{figure}
    \includegraphics[width=\figurewidth]{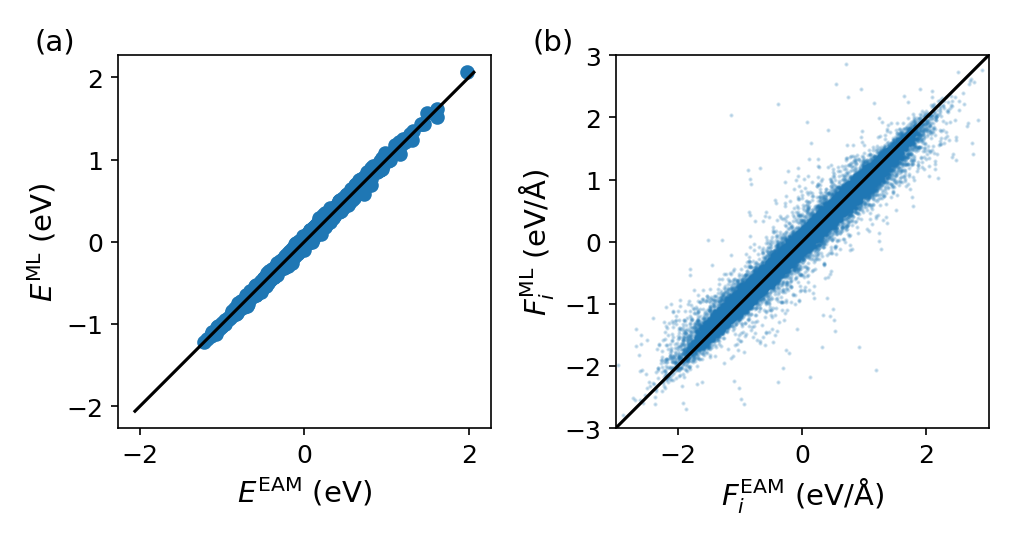}
    \caption{ML model predictions on the 2000 Cu clusters. \\a) ML predicted energies energies compared to EAM values. b) Atomic force components compared to EAM forces.
    }
    \label{fig:fig-6}
\end{figure}
We assessed the accuracy of the Cu cluster model fit by predicting the total potential energies of 2 000 Cu test clusters and comparing to EAM energies. Fig. \ref{fig:fig-6}a shows the results of this comparison. The absolute error of the ML model predictions is only \SI{0.43}{meV/atom} on average. Next, we tested the force prediction accuracy of the ML model by comparing the ML predicted atomic forces of the same 2 000 Cu clusters to the EAM forces. Fig. \ref{fig:fig-6}b depicts atomic force components computed with both methods. The mean absolute error of the predictions is \SI{43}{meV/\text{\AA}}.

By combining the SOAP descriptor with a simple neural network architecture, we were able achieve good accuracy in both energy and force prediction. Here, we trained the ML model on classical MD data for demonstrative purposes, and no speed up was achieved. The same methodology, however, could be applied on data from a more accurate method, such as DFT, in which it would greatly accelerate simulations.

\section{\label{sec:conclusions}Conclusions}
We have presented an update of the DScribe package. We have introduced a new structural representation, expanding DScribe's descriptor selection. Additionally, we have extended the capabilities of DScribe by implementing descriptor derivatives. We have compared the accuracy of our numerical derivative implementation to analytical MBTR and SOAP derivatives, and observe relative errors of less than $10^{-6}.$ We have also demonstrated the effectiveness of the analytical derivative implementations through two machine learning tasks involving force prediction and structure optimization. Our results show that our derivative implementations are accurate and effective, and we believe that the new version of DScribe will be a valuable tool for researchers applying machine learning to materials science problems.

\section*{Supplementary material}
See supplementary material for the detailed derivation of the analytical MBTR and SOAP derivatives.

\begin{acknowledgments}
We acknowledge funding from the European Union’s Horizon program under the grant agreement No 951786 and the Academy of Finland through project No 334532 and the Center of Excellence Virtual Laboratory for Molecular Level Atmospheric Transformations (VILMA; project No 346377). We further acknowledge CSC-IT Center for Science, Finland, and the Aalto Science-IT project.
\end{acknowledgments}


\section*{References}
\FloatBarrier
\bibliography{ref}


\clearpage
\onecolumngrid

\begin{center}
	\textbf{\large Supplementary Material: \\
		Updates to the DScribe Library: New Descriptors and Derivatives} \\
	\vspace{1em}
	Jarno Laakso,\textsuperscript{1} Lauri Himanen,\textsuperscript{1} Henrietta Homm,\textsuperscript{1} Eiaki V. Morooka,\textsuperscript{1} Marc O. J. J\"ager,\textsuperscript{1} Milica Todorovi\'c,\textsuperscript{2} and Patrick Rinke\textsuperscript{1} \\
	\textsuperscript{1)}\textit{Department of Applied Physics, Aalto University, P.O. Box 11100, 00076 Aalto, Finland} \\
	\textsuperscript{2)}\textit{Department of Mechanical and Materials Engineering, University of Turku, FI-20014 Turku, Finland}
\end{center}

\setcounter{equation}{0}
\setcounter{section}{0}
\setcounter{figure}{0}
\setcounter{table}{0}
\setcounter{page}{1}
\makeatletter

\renewcommand{\theequation}{S\arabic{equation}}
\renewcommand{\thefigure}{S\arabic{figure}}
\renewcommand{\thetable}{S\Roman{table}}
\renewcommand{\bibnumfmt}[1]{[S#1]}
\renewcommand{\citenumfont}[1]{S#1}

\newcommand{\xminmm}{x_{\text{min}}}
\newcommand{\xmaxmm}{x_{\text{max}}}
\newcommand{\ngridmm}{N_{\text{grid}}}
\newcommand{\wcutmm}{w_{\text{cutoff}}}
\newcommand{\rcutmm}{r_{\text{cutoff}}}

\def\xmin/{$\xminmm$}
\def\xmax/{$\xmaxmm$}
\def\ngrid/{$\ngridmm$}
\def\wcut/{$\wcutmm$}
\def\rcut/{$\rcutmm$}

\renewcommand{\thepage}{S\arabic{page}}
\renewcommand{\thesection}{S\arabic{section}}
\renewcommand{\thetable}{S\arabic{table}}
\renewcommand{\thefigure}{S\arabic{figure}}
\newcommand{\ra}[1]{\renewcommand{\arraystretch}{#1}}

\section{Notation}
We adopt a notation where the vector between the coordinates of atoms $l$ and $m$ is
\begin{align}
\vect{R}_l-\vect{R}_m \equiv \vect{r}_{lm}\\
|\vect{r}_{lm}| \equiv r_{lm}
\end{align}

\section{Many-body Tensor Representation (MBTR)}

\subsection{Definition}
MBTR describes atomic geometries as distributions of small structural motifs that come in different sizes $k$. The \texttt{dscribe} implementation has terms up to $k=3.$ The $k=1$ term considers single atoms and relates to the elemental content of the atomic structure. The $k=2$ term considers atom pairs and their distances. The $k=3$ term corresponds to atom triplets and the angles between them. Each term $k$ of the representation is a collection of functions that correspond to every possible combination of elements that can appear in a group of $k$ atoms:
\begin{align}
\text{MBTR}_{1}^{Z_1}(x) =& \sum_{l}^{|Z_1|}w_1^{l}d_1^{l}(x)\label{eqs:mbtr_1}\\
\text{MBTR}_{2}^{Z_1,Z_2}(x) =& \sum_{l}^{|Z_1|}\sum_{m}^{|Z_2|}w_2^{l,m}d_2^{l,m}(x)\label{eqs:mbtr_2}\\
\text{MBTR}_{3}^{Z_1,Z_2,Z_3}(x) =& \sum_{l}^{|Z_1|}\sum_{m}^{|Z_2|}\sum_{n}^{|Z_3|}w_3^{l,m,n}d_3^{l,m,n}(x),\label{eqs:mbtr_3}
\end{align}
where the sums run over all the atom groups that have elements $Z_1,$ $Z_2,$ and $Z_3.$ Each function is a weighted sum of normal distributions $d_k(x):$
\begin{align}
d_1^{l}(x) =& \frac{1}{\sigma_1 \sqrt{2\pi}}e^{-\frac{(x-g_1^{l})^2}{2\sigma_1^2}}\label{eqs:d_1}\\
d_2^{l,m}(x) =& \frac{1}{\sigma_2 \sqrt{2\pi}}e^{-\frac{(x-g_2^{l,m})^2}{2\sigma_2^2}}\label{eqs:d_2}\\
d_3^{l,m,n}(x) =& \frac{1}{\sigma_3 \sqrt{2\pi}}e^{-\frac{(x-g_3^{l,m,n})^2}{2\sigma_3^2}},\label{eqs:d_3}
\end{align}
where $\sigma_k$ determine the widths of the distributions. $g_k$ are so called geometry functions that map groups of atoms to scalar values. We use:
\begin{align}
g_1^{l} =& Z_l\label{eqs:g_1}\\
g_{2,\text{dist.}}^{l,m} = r_{lm}\hspace{3mm}\text{or}&\hspace{3mm} g_{2,\text{inv.dist.}}^{l,m} = \frac{1}{r_{lm}}\label{eqs:g_2}\\
g_3^{l,m,n} =&  \cos(\angle(\vect{r}_{lm},\vect{r}_{nm})) =  \frac{r_{ml}^2+r_{mn}^2-r_{ln}^2}{2r_{ml}r_{mn}}\label{eqs:g_3}
\end{align}
$w_k$ in (\ref{eqs:mbtr_1}-\ref{eqs:mbtr_3}) are weighting functions that guarantee that the sums converge when calculating the MBTR representations of periodic structures. We use:
\begin{align}
w_1^{l}=&1\label{eqs:w_1}\\
w_{2,\text{inv.square}}^{l,m} = \frac{1}{r_{lm}^2} \hspace{3mm}\text{or}&\hspace{3mm} w_{2,\text{exp}}^{l,m}=e^{-s_2 r_{lm}}\label{eqs:w_2}\\
w_3^{l,m,n}=&e^{-s_3(r_{lm}+r_{mn}+r_{nl})}\label{eqs:w_3}
\end{align}

The MBTR vectors are obtained by evaluating the functions (\ref{eqs:mbtr_1}-\ref{eqs:mbtr_3}) at discreet grid points
\begin{align}
\vect{x} = (\xminmm,\xminmm+\Delta x,\, ...\, ,\xmaxmm),
\end{align}
where \xmin/ and \xmax/ are the limits of the grid and $\Delta x$ is the spacing between two grid points
\begin{align}
\Delta x = \frac{x_{\max}-x_{\min}}{\ngridmm-1},
\end{align}
\ngrid/ is the number of grid points. This can be done by discretizing the distribution function $d_k(x)$ into a vector $\vect{d}_k: (\vect{d}_k)_i = d_k(\vect{x_i}).$ However, instead of evaluating $d_k(x)$ directly at the grid points, we obtain the function values by differentiating numerically its cumulative distribution
\begin{align}
D_k(x) = \int_{-\infty}^x d_k(x^{\prime})dx^{\prime}=\frac{1}{2}\left[ 1+\text{erf}\left(\frac{x-g_k}{\sigma\sqrt{2}}\right) \right]
\end{align}
through
\begin{align}
d_k(x) = \frac{D_k(x+\frac{\Delta x}{2})-D_k(x-\frac{\Delta x}{2})}{\Delta x}.
\label{eqs:CDF_to_d}
\end{align}
Having obtained $\vect{d}_k,$ the MBTR vectors are given by
\begin{align}
\vect{M}_1^{Z_1} =& \sum_{l}^{|Z_1|}w_1^{l,m}\vect{d}_1^{l}\label{eqs:vec_mbtr_1}\\
\vect{M}_2^{Z_1,Z_2} =& \sum_{l}^{|Z_1|}\sum_{m}^{|Z_2|}w_2^{l,m}\vect{d}_2^{l,m}\label{eqs:vec_mbtr_2}\\
\vect{M}_3^{Z_1,Z_2,Z_3} =& \sum_{l}^{|Z_1|}\sum_{m}^{|Z_2|}\sum_{n}^{|Z_3|}w_3^{l,m,n}\vect{d}_3^{l,m,n}\label{eqs:vec_mbtr_3}
\end{align}
The elemental contributions are concatenated into $\vect{M}_1,$ $\vect{M}_2,$ and $\vect{M}_3,$ which are then concatenated further to obtain the full vector representation:
\begin{align}
\vect{M} = \vect{M}_1^\frown\vect{M}_2^\frown\vect{M}_3.
\end{align}

\subsection{Derivatives}
Let us now compute the derivatives of the MBTR vector representation with respect to the position of atom $a.$ In order to derive a general result for $\nabla_a \vect{M}_k$ for any $k$, we first write down the gradient of a distribution function
\begin{align}
\nabla_a d_k(x) =& \nabla_a \left( \frac{1}{\sigma_2 \sqrt{2\pi}}e^{-\frac{(x-g_k)^2}{2\sigma_k^2}} \right) \nonumber\\
=& \frac{1}{\sigma_2 \sqrt{2\pi}} e^{-\frac{(x-g_k)^2}{2\sigma_k^2}} \nabla_a \left( -\frac{(x-g_k)^2}{2\sigma_k^2} \right)\nonumber\\
=& d_k(x) \frac{1}{\sigma_k^2} (x-g_k) \nabla_a g_k
\label{eqs:d_d}
\end{align}
and the weighted distribution
\begin{align}
\nabla_a \left(w_kd_k(x)\right) =& d_k\nabla_a w_k + w_k\nabla_a d_k(x)\nonumber\\
=& d_k(x)\nabla_a w_k + w_k d_k(x) \frac{1}{\sigma_k^2} (x-g_k) \nabla_a g_k\nonumber\\
=& d_k(x)\left(\nabla_a w_k-w_k\frac{1}{\sigma_k^2}g_k\nabla_a g_k\right) + xd_k(x)\left(w_k\frac{1}{\sigma_k^2}\nabla_a g_k\right)
\label{eqs:wd_d}
\end{align}
Here we have separated the two differently shaped distributions $d_k(x)$ and $xd_k(x).$ While discretizing the representation, both of them need to be estimated through their cumulative distributions. This is already done for the $d_k(x)$ when calculating the representation itself. We give the second distribution its own name: $c_k(x):= xd_k(x).$ Its cumulative distribution is 
\begin{align}
C_k(x) =& \int_{-\infty}^x x^{\prime} d_k(x^{\prime})dx^{\prime}\nonumber\\
=& \frac{1}{\sigma_k\sqrt{2\pi}} \int_{-\infty}^x x^{\prime} e^{-\frac{(x^{\prime}-g_k)^2}{2\sigma_k^2}}dx^{\prime}\nonumber\\
=& g_k\frac{1}{2}\left[ 1+\text{erf}\left(\frac{x-g_k}{\sigma_k\sqrt{2}}\right)\right] - \frac{\sigma_k}{\sqrt{2\pi}}\left[ e^{-\frac{(x-g_k)^2}{2\sigma_k^2}}-1\right]\nonumber\\
=& g_kD_k(x) - \frac{\sigma_k}{\sqrt{2\pi}}\left[ e^{-\frac{(x-g_k)^2}{2\sigma_k^2}}-1\right]
\end{align}
$C_k(x)$ can be used to estimate $c_k(x)$ at discrete $x-$values similarly to (\ref{eqs:CDF_to_d}) in order to obtain vectors $\vect{c}_k$. The derivative of an MBTR vector is given by
\begin{align}
\nabla_a \vect{M}_k =& \nabla_a \sum w_k\vect{d}_k(x)\nonumber\\
=& \sum \nabla_a\left(w_k\vect{d}_k(x)\right)\nonumber\\
=& \sum \vect{d}_k\left(\nabla_a w_k-w_k\frac{1}{\sigma_k^2}g_2^{l,m}\nabla_a g_k\right) + \vect{c}_k\left(w_2^{l,m}\frac{1}{\sigma_2^2}\nabla_a g_k\right),
\label{eqs:mbtr_d}
\end{align}

Now we can easily derive the gradients for different $k$. $\nabla_a \vect{M}_1=0$ because it does not depend on the atomic positions. $\nabla_a  \vect{M}_2$ and $\nabla_a  \vect{M}_3$ are nonzero. We write down some auxiliary results:
\begin{align}
\nabla_a r_{lm} =& (\delta_{al}-\delta_{am})\frac{\vect{r}_{lm}}{r_{lm}} = (\delta_{al}-\delta_{am})\vect{\hat{r}}_{lm}\label{eqs:r_d_atom}\\
\nabla_a r_{lm}^2 =& 2r_{lm}\nabla_a r_{lm} = (\delta_{al}-\delta_{am})2\vect{r}_{lm}\label{eqs:r2_d_atom}
\end{align}
Using (\ref{eqs:mbtr_d}), the gradients of $\vect{M}_2$ are
\begin{align}
\nabla_a \vect{M}_2^{Z_1,Z_2} =&  \sum_{l}^{|Z_1|}\sum_{m}^{|Z_2|}\left[\vect{d}_2^{l,m}\left(\nabla_a w_2^{l,m}-w_2^{l,m}\frac{1}{\sigma_2^2}g_2^{l,m}\nabla_ag_2^{l,m}\right) + \vect{c}_2^{l,m}\left(w_2^{l,m}\frac{1}{\sigma_2^2}\nabla_a g_2^{l,m}\right)\right],
\label{eqs:mbtr_2_d_atoms}
\end{align}
where the gradients for the different $g_2^{l,m}$ and $w_2^{l,m}$ options can be solved using (\ref{eqs:r_d_atom}) and (\ref{eqs:r2_d_atom}):
\begin{align}
\nabla_a g_{2,\text{dist.}}^{l,m} =& \nabla_a r_{lm} = (\delta_{al}-\delta_{am})\vect{\hat{r}}_{lm}\label{eqs:g_2_d_atoms_dist}\\
\nabla_a g_{2,\text{inv.dist.}}^{l,m} =& \nabla_a \frac{1}{r_{lm}} = -\frac{1}{r_{lm}^2} \nabla_a r_{lm} = -(\delta_{al}-\delta_{am})\frac{\vect{r}_{lm}}{r_{lm}^3}\label{eqs:g_2_d_atoms_inv_dist}\\
\nabla_a w_{2,\text{inv.square}}^{l,m} =& \nabla_a \frac{1}{r_{lm}^2} = -\frac{2}{r_{lm}^3} \nabla_a r_{lm} = -2(\delta_{al}-\delta_{am})\frac{\vect{r}_{lm}}{r_{lm}^4}\label{eqs:w_2_d_atoms_inv_square}\\
\nabla_a w_{2,\text{exp}}^{l,m} =& \nabla_a e^{-s_2r_{lm}} = -e^{-s_2r_{lm}} s_2 \nabla_a r_{lm} = -(\delta_{al}-\delta_{am})w_{2,\text{exp}}^{l,m}s_2\vect{\hat{r}}_{lm}.\label{eqs:w_2_d_atoms_exp}
\end{align}
Similarly for $\vect{M}_3:$
\begin{align}
\nabla_a \vect{M}_3^{Z_1,Z_2,Z_3} =&  \sum_{l}^{|Z_1|}\sum_{m}^{|Z_2|}\sum_{n}^{|Z_3|}\left[\vect{d}_3^{l,m,n}\left(\nabla_a w_3^{l,m,n}-w_3^{l,m,n}\frac{1}{\sigma_3^2}g_3^{l,m,n}\nabla_ag_3^{l,m,n}\right) + \vect{c}_3^{l,m,n}\left(w_3^{l,m,n}\frac{1}{\sigma_3^2}\nabla_a g_3^{l,m,n}\right)\right].
\label{eqs:mbtr_3_d_atoms}
\end{align}
The gradient of the geometry function is 
\begin{align}
\nabla_a(g_3^{l,m,n}) =& \nabla_a\left( \frac{r_{ml}^2+r_{mn}^2-r_{ln}^2}{2r_{ml}r_{mn}} \right)\nonumber\\
=& \frac{r_{ml}r_{mn}\nabla_a(r_{ml}^2+r_{mn}^2-r_{ln}^2) - (r_{ml}^2+r_{mn}^2-r_{ln}^2)\nabla_a(r_{ml}r_{mn})}{2r_{ml}^2r_{mn}^2},\label{eqs:g_3_d_atoms}
\end{align}
where
\begin{align}
\nabla_a(r_{ml}^2+r_{mn}^2-r_{ln}^2) = 2\left[ (\delta_{am}-\delta_{al})\vect{\hat{r}}_{ml} + (\delta_{am}-\delta_{an})\vect{\hat{r}}_{mn} - (\delta_{al}-\delta_{an})\vect{\hat{r}}_{ln} \right]
\end{align}
and
\begin{align}
\nabla_a(r_{ml}r_{mn}) = (\delta_{am}-\delta_{an})r_{ml}\vect{\hat{r}}_{mn} + (\delta_{am}-\delta_{al})r_{mn}\vect{\hat{r}}_{ml}.
\end{align}
The gradient of the weight function is
\begin{align}
\nabla_aw_3^{l,m,n} =& \nabla_a e^{-s_3(r_{lm}+r_{mn}+r_{nl})}\nonumber\\
=& -s_3 w_3^{l,m,n}\nabla_a(r_{lm}+r_{mn}+r_{nl})\nonumber\\
=& -s_3 w_3^{l,m,n}\left[ (\delta_{al}-\delta_{am})\vect{\hat{r}}_{lm} + (\delta_{am}-\delta_{an})\vect{\hat{r}}_{mn} + (\delta_{an}-\delta_{al})\vect{\hat{r}}_{nl} \right].\label{eqs:w_3_d_atoms}
\end{align}

\section{Smooth Overlap of Atomic Positions (SOAP) with Gaussian type orbital (GTO) radial basis}

\subsection{Definition}
The final output from our SOAP implementation is the partial power spectra
vector $\mathbf{p}$ where the individual vector elements are defined as
\begin{align}
p^{Z_1,Z_2}_{nn'l}
&= \pi \sqrt{\frac{8}{2l+1}} \sum_m \left(c_{nlm}^{Z_1}\right)^*c^{Z_2}_{n'lm} \label{eqs:power_spectrum_complex_1} \\
&= \pi \sqrt{\frac{8}{2l+1}} \sum_m \left(\sum_j^{\lvert Z_1 \rvert} c^j_{nlm}\right)\left(\sum_k^{\lvert Z_2 \rvert} c^k_{n'lm}\right)  \label{eqs:power_spectrum_complex_2}
\end{align}
Because we are using real spherical harmonics, the complex conjugation in
(\ref{eqs:power_spectrum_complex_1}) can be omitted. The summation for
$j$ and $k$ run over atoms with the atomic number $Z_1$ and $Z_2$ respectively
and the function $c^i_{nlm}$ is defined as
\begin{align}
c^i_{nlm}
&= \iiint_{\mathcal{R}^3}\mathrm{d}V g_{nl}(r)\rho(\bm{r}, \bm{r}_i)Y_{lm}(\theta, \phi) \label{eqs:coefficient_atom}
\end{align}
The real spherical harmonics are defined in spherical coordinates as
\begin{equation}
Y_{lm}(\theta, \phi) = 
\begin{cases}
\sqrt{2} (-1)^m \text{Im}[ Y_l^{\lvert m \lvert}(\theta, \phi)] & \text{if }m < 0 \\
Y_l^0 & \text{if }m = 0 \\
\sqrt{2} (-1)^m \text{Re}[Y_l^{m}(\theta, \phi)] & \text{if }m > 0 \\
\end{cases}
\end{equation}
where $Y_l^m$ corresponds to the complex orthonormalized spherical harmonics defined as
\begin{equation}
Y_l^m(\theta, \phi) = \sqrt{\frac{(2l+1)}{4\pi}\frac{(l-m)!}{(l+m)!}}P_l^m(\cos\theta)e^{im\phi}
\end{equation}
and $P_l^m$ are the associated Legendre polynomials. We will also be using the definition and properties of real regular solid harmonics $R_{l}^{m}$
which are defined as
\begin{equation}
R_{l}^{m}(\bm{r}) = \sqrt{\frac{4\pi}{2l+1}} r^l Y_{lm}(\bm{r})
\end{equation}
These real regular solid harmonics, expressed in Cartesian coordinates, are real-valued homogeneous polynomials of order $l$ in $x$, $y$, $z$. Using the definition of the GTO radial basis $g_{nl}(r)$
\begin{align}
g_{nl}(r) &= \sum_{n'=1}^{n_\mathrm{max}}\,\beta_{nn'l} r^l e^{-\alpha_{n'l}r^2}
\label{eqs:gto}
\end{align}
and the definition of the Gaussian atomic density $\rho(\bm{r}, \bm{r}_i)$
\begin{align}
\rho(\bm{r}, \bm{r}_i)
&= e^{-\frac{1}{2 \sigma^2} ( \bm{r} -\bm{r}_i ) ^2} \\
&= e^{-s ( \bm{r} -\bm{r}_i ) ^2} \label{eqs:atomic_density}
\end{align}
one can simplify (\ref{eqs:coefficient_atom}) as follows
\begin{align}
c^i_{nlm}
&= \iiint_{\mathcal{R}^3} \mathrm{d}V \sum_{n'=1}^{n_\mathrm{max}} \beta_{nn'l} r^l e^{-\alpha_{n'l} r^2} e^{-s(\bm{r} - \bm{r_i})^2} Y_{lm}(\bm{r}) \\
&= \iiint_{\mathcal{R}^3} \mathrm{d}V \sum_{n'=1}^{n_\mathrm{max}} \beta_{nn'l} e^{-\alpha_{n'l} r^2 -s(\bm{r} - \bm{r_i})^2} r^l Y_{lm}(\bm{r}) \label{eqs:uncompleted} \\
&= \iiint_{\mathcal{R}^3} \mathrm{d}V \sum_{n'=1}^{n_\mathrm{max}} \beta_{nn'l} e^{-\frac{s \alpha_{n'l}}{s + \alpha_{n'l}}r_i^2 - \left( s + \alpha_{n'l}\right)\left(\bm{r} - \frac{s\bm{r}_i}{s + \alpha_{n'l}}\right)^2} r^l Y_{lm}(\bm{r}) \label{eqs:completed} \\
&= \sum_{n'=1}^{n_\mathrm{max}} \beta_{nn'l} e^{-\frac{s\alpha_{n'l}}{s+\alpha_{n'l}} r^2_i} \iiint_{\mathcal{R}^3} \mathrm{d}V e^{- (s+\alpha_{n'l})(\bm{r} - \frac{s\bm{r_i}}{s+\alpha_{n'l}})^2} r^l Y_{lm}(\bm{r}) \label{eqs:before_transform} \\
&= \sum_{n'=1}^{n_\mathrm{max}} \beta_{nn'l} e^{-\frac{s\alpha_{n'l}}{s+\alpha_{n'l}} r^2_i} \iiint_{\mathcal{R}^3} \mathrm{d}V e^{- (s+\alpha_{n'l})r^2} r^l Y_{lm}(\bm{r} + \frac{s\bm{r}_i}{s+\alpha_{n'l}}) \label{eqs:after_transform} \\
&= \sum_{n'=1}^{n_\mathrm{max}} \beta_{nn'l} e^{-\frac{s\alpha_{n'l}}{s+\alpha_{n'l}} r^2_i} \sqrt{\frac{2l+1}{4\pi}} \int_{0}^{\infty} r^2 e^{- (s+\alpha_{n'l})r^2} \int_0^{2\pi} \int_0^{\pi} \sin{\theta} R_{l}^{m}(\bm{r} + \frac{s\bm{r}_i}{s+\alpha_{n'l}}) \mathrm{d}\theta \mathrm{d}\phi \mathrm{d}r \label{eqs:before_addition} \\
\begin{split}
&= \sum_{n'=1}^{n_\mathrm{max}} \beta_{nn'l} e^{-\frac{s\alpha_{n'l}}{s+\alpha_{n'l}} r^2_i} \sqrt{\frac{2l+1}{4\pi}} \int_{0}^{\infty}r^2 e^{- (s+\alpha_{n'l})r^2} \int_0^{2\pi} \int_0^{\pi} \sum_{\lambda=0}^\ell\binom{2\ell}{2\lambda}^{1/2} \sum_{\mu=-\lambda}^\lambda R^\mu_{\lambda}(\mathbf{r}) R^{m-\mu}_{\ell-\lambda}(\frac{s\bm{r}_i}{s+\alpha_{n'l}})\; \\
&\qquad \langle \lambda, \mu; \ell-\lambda, m-\mu| \ell m \rangle \mathrm{d}\theta \mathrm{d}\phi \mathrm{d}r \label{eqs:addition_theorem}
\end{split}
\\[2ex]
&= \sum_{n'=1}^{n_\mathrm{max}} \beta_{nn'l} e^{-\frac{s\alpha_{n'l}}{s+\alpha_{n'l}} r^2_i} 4\pi \sqrt{\frac{2l+1}{4\pi}} R_{l}^{m}(\frac{s\bm{r}_i}{s+\alpha_{n'l}}) \int_{0}^{\infty} r^2 e^{- (s+\alpha_{n'l})r^2} \mathrm{d}r \label{eqs:addition_integrated} \\
&= \frac{\pi^{3/2}}{(s+\alpha_{n'l})^{3/2}}\sum_{n'=1}^{n_\mathrm{max}} \beta_{nn'l} e^{-\frac{s\alpha_{n'l}}{s+\alpha_{n'l}} r^2_i} (\frac{s r_i}{s+\alpha_{n'l}})^l Y_{lm}(\frac{s\bm{r}_i}{s+\alpha_{n'l}}) \label{eqs:scaled} \\
&= \frac{\pi^{3/2} s^{l}}{(s+\alpha_{n'l})^{3/2 + l}} \sum_{n'=1}^{n_\mathrm{max}} \beta_{nn'l} e^{-\frac{s\alpha_{n'l}}{s+\alpha_{n'l}} r^2_i} r_i^l Y_{lm}(\bm{r}_i) \label{eqs:unscaled}
\end{align}
In the transition from (\ref{eqs:uncompleted}) to (\ref{eqs:completed}) we have completed a square in the exponent in order to obtain
a form containing $(\bm{r} - \frac{s \bm{r}_i}{s+\alpha_{n'l}})^2$. This allows us to perform the substitution $\bm{r} \rightarrow
\bm{r} + \frac{s\bm{r}_i}{s+\alpha_{n'l}}$ when transitioning from (\ref{eqs:before_transform}) to (\ref{eqs:after_transform}), which
does not affect the differential or the limits at infinity. In the transition from (\ref{eqs:before_addition}) to (\ref{eqs:addition_theorem})
we have utilized the addition theorem for solid harmonics, where $\langle \lambda, \mu; \ell-\lambda, m-\mu| \ell m \rangle$ are Clebsch–Gordan coefficients. Upon integrating over the angular coordinates, only the term with $\lambda = \mu = 0$ remains,
and the remaining radial integral contains a Gaussian integral with a polynomial prefactor that has a known analytical form. In the transition
from (\ref{eqs:scaled}) to (\ref{eqs:unscaled}) we have used the fact that scaling the radial variable does not affect the spherical harmonics, i.e. $Y_{lm}(a\bm{r}) = Y_{lm}(\bm{r})$.

\subsection{Derivatives}
The derivative of the power spectrum (\ref{eqs:power_spectrum_complex_2}) with respect to the Cartesian coordinate $x_i$ of atom $i$ is given by
\begin{align}
\frac{\partial p^{Z_1,Z_2}_{nn'l}}{ \partial x_i}
&= \pi \sqrt{\frac{8}{2l+1}} \sum_m \frac{\partial}{ \partial x_i} \left[ \left(\sum_j^{\lvert Z_1 \rvert} c^j_{nlm}\right)\left(\sum_k^{\lvert Z_2 \rvert} c^k_{n'lm}\right) \right] \\
&= \pi \sqrt{\frac{8}{2l+1}} \sum_m \left[ \frac{\partial}{ \partial x_i} \left( \sum_j^{\lvert Z_1 \rvert} c^j_{nlm} \right) \sum_k^{\lvert Z_2 \rvert} c^k_{n'lm} + \sum_j^{\lvert Z_1 \rvert} c^j_{nlm} \frac{\partial}{ \partial x_i} \left( \sum_k^{\lvert Z_2 \rvert} c^k_{n'lm} \right) \right] \\
&= \pi \sqrt{\frac{8}{2l+1}} \sum_m \left[ \left( \sum_j^{\lvert Z_1 \rvert} \frac{\partial c^j_{nlm}}{ \partial x_i} \right) \sum_k^{\lvert Z_2 \rvert} c^k_{n'lm} + \sum_j^{\lvert Z_1 \rvert} c^j_{nlm} \left( \sum_k^{\lvert Z_2 \rvert} \frac{\partial c^k_{n'lm}}{ \partial x_i} \right) \right] \label{eqs:pnnl_derivative}
\end{align}
From this form, it can be seen that the derivative calculation mainly consists of the derivatives $\frac{\partial c^i_{nlm}}{ \partial x_i}$. Using the definition from (\ref{eqs:unscaled}), the required derivatives $\frac{\partial c^i_{nlm}}{ \partial x_i}$ can be written as
\begin{align}
\frac{\partial c^i_{nlm}}{ \partial x_i} &= \frac{\pi^{3/2} s^{l}}{(s+\alpha_{n'l})^{3/2 + l}}\sum_{n'=1}^{n_\mathrm{max}} \beta_{nn'} e^{-\frac{s\alpha_{n'l}}{s+\alpha_{n'l}} r^2_i}
\left[ \frac{\partial r_i^l Y_{lm}(\bm{r}_i)}{\partial x_i} - \frac{2\alpha_{n'l}x_i}{s+\alpha_{n'l} } r_i^l Y_{lm}(\bm{r}_i) \right] \label{eqs:cln_derivative}
\end{align}
As $r_i^l Y_{lm}(\bm{r}_i)$ is a polynomial function of $x_i$, $y_i$ and $z_i$ it can be differentiated relatively easily. These derivatives can be inserted into (\ref{eqs:cln_derivative}), which can be used in (\ref{eqs:pnnl_derivative}) to obtain the final analytical formulas for the derivatives. Identical analysis can be carried out also for $\frac{\partial p^{Z_1,Z_2}_{nn'l}}{ \partial y_i}$ and $\frac{\partial p^{Z_1,Z_2}_{nn'l}}{ \partial z_i}$.

\end{document}